\documentclass[aps,pra,superscriptaddress,twocolumn]{revtex4-1}

\usepackage{graphicx} 
\usepackage{amsmath}
 
\newcommand{\beq}{\begin{equation}}
\newcommand{\enq}{\end{equation}}
\newcommand{\bea}{\begin{eqnarray}}
\newcommand{\ena}{\end{eqnarray}}
\newcommand{\ah}{\hat{a}}
\newcommand{\ad}{\hat{a}^{\dag}}
\begin{document}

\title{Mott insulator dynamics}
\author{Emil Lundh}
\affiliation{Department of Physics, Ume{\aa} University, 901 87 Ume\aa, Sweden}
\date{\today}
\begin{abstract}
The hydrodynamics of a lattice Bose gas in a time-dependent 
external potential is 
studied in a mean-field approximation. The conditions under which a 
Mott insulating region can melt, and the local density adjust to the 
new potential, are determined. 
In the case of a suddenly switched potential, it is found that the Mott 
insulator stays insulating and the density will not adjust if the switch 
is too abrupt. 
This comes about because too rapid currents result in Bloch oscillation-type 
current reversals. 
For a stirrer moved through a Mott insulating cloud, it is seen that 
only if the stirrer starts in a superfluid region and the 
velocity is comparable to the time scale set by the tunneling, 
will the Mott insulator be affected. 
\end{abstract}
\maketitle

\section{Introduction}
\label{sec:intro}
When Greiner {\it et al.} put a system of bosonic atoms in an 
optical lattice and demonstrated the Mott transition \cite{greiner2002}, 
research on the physics of strongly correlated systems entered a new phase.
Cold atoms in optical lattices offer control and detection techniques 
unthinkable in a condensed matter system. 
The versatility of optical lattices have teased the imagination of 
theorists, and copious amounts of papers suggesting exotic types of 
quantum phase transitions have been published in the past ten years
\cite{bloch2008}. However, almost all of the suggested quantum 
phases have proven impossible to realize experimentally at the present 
for various technical reasons. For example, many rely on spin 
ordering, which requires temperatures far below what is possible at 
the present (see, however, the ingenious experiment by Simon {\it et al.}
using occupation as an effective spin degree of freedom to realize a quantum Ising model \cite{simon2011}). 

While the quest for achieving lower temperatures is on \cite{ho2007}, 
there is still lots of interesting science to be done with the 
relatively simple Bose-Hubbard model on a square lattice. With the 
possibilities of monitoring directly the local density and low-order 
correlation functions on the one hand 
\cite{folling2006,scherson2010}, and the 
possibility to manipulate external potentials in real time on the 
other hand \cite{schori2004}, the idea to study the real-time 
dynamics of a strongly correlated system suggests itself. 
In particular, these systems are by default trapped in inhomogeneous 
potentials and therefore they contain finite regions of superfluid and 
Mott insulator sitting side by side. With current techniques, one 
can investigate how the interfaces between such regions behave in 
real time under various kinds of perturbation. 

The majority of studies of dynamics of lattice Bose systems have been 
concerned with quenches \cite{clark2004}, Bloch oscillations 
\cite{sachdev2002}, and collective modes in traps 
\cite{pupillo2003,lundh2004a,lundh2004b}.
These aspects of trapped lattice bosons are by now well understood. 
Worth mentioning is also an important study of the critical current in 
interacting lattice 
Bose systems \cite{polkovnikov2005}. 
Concerning the macroscopic transport of matter and redistribution 
between quantum phases due to potentials and currents -- 
i.e., the hydrodynamics -- the literature 
is less complete. Natu {\it et al.} study the redistribution of matter 
within an optical lattice following a quench \cite{natu2011}. 
Fischer {\it et al.} were concerned with the velocity of a moving 
superfluid-Mott insulator interface \cite{fischer2008}. 
Karlsson {\it et al.} study the behavior of a trapped partly Mott 
insulating system after turn-off of the trap \cite{karlsson2011}, and 
Snoek and Hofstetter study the dynamics upon displacement of the trap 
\cite{snoek2007}; this paper will be further discussed below.

The literature survey above indicates that there is need for a more 
exhaustive understanding of how Mott insulators react to macroscopic 
currents and potential gradients, which is the subject of this paper. 
It will provide two sets of numerical examples of how a Bose-Hubbard 
system containing coexisting Mott and superfluid regions evolves 
in time following a perturbation in the potential. 
In order to do this, we evolve the Bose-Hubbard Hamiltonian in time 
within the mean-field Gutzwiller approximation. 
In Sec.\ \ref{sec:model} the governing equations of motion are put up. 
In Sec.\ \ref{sec:gaussian}, I study the in- and outflow to or from a 
Mott insulator following the onset of a potential gradient. 
Sec.\ \ref{sec:spoon} discusses the perturbation of a Mott 
insulator using a localized stirrer. In Sec.\ \ref{sec:conclusion} 
I summarize and conclude.

\section{Lattice Bose gas} 
\label{sec:model}

We study a system of zero-temperature bosons hopping on a 
lattice and in addition subject to a more slowly varying external 
potential. 
The many-body Bose-Hubbard Hamiltonian is 
\bea
H &=& 
- J \sum_{<jj'>}\ad_{j} \ah_{j'} 
+ \frac{U}{2} \sum_{j}\ad_{j}\ad_{j} \ah_{j} \ah_{j} 
\nonumber\\
&+& \sum_{j} (V(r_j)-\mu) \ad_{j}\ah_{j}.
\label{hamiltonian}
\ena
Here, $J$ is the tunneling matrix element, $U$ is the interaction 
parameter and $\mu$ is the chemical potential. 
$V(r_j)$ is the external potential 
and the index $j$ runs over the lattice sites. $r_j$ is the 
spatial coordinate of the $j$:th lattice site.
The sum subscripted $<j,j'>$ runs over nearest neighbors. 
We work in units in which $\hbar=1$ and the lattice constant is 
also unity; hence, both frequencies and velocities can be measured in 
units of $U$. 
This paper will be concerned with 
a two-dimensional square lattice. 
However, note that the mean-field approximation that will be 
used here is most accurate in higher dimensions, so the 
results obtained are best seen as qualitatively describing 3D 
physics. 

The Gutzwiller approximation is based on a mean-field ansatz for 
the many-body state \cite{jaksch1998}, 
\beq
|\psi_G(t)\rangle = \prod_j  |\phi_{j}(t)\rangle.
\label{gutzwiller} 
\enq
In the numerical computations, the on-site states are expanded 
in a local Fock basis with an upper cutoff $n_{\rm max}$, 
\beq
|\phi_{j}(t)\rangle = \sum_{n=0}^{n_{\rm max}} C_{j,n}(t) |n\rangle_{r}. 
\enq
In order to calculate the ground state, the Hamiltonian 
(\ref{hamiltonian}) is minimized with respect to the complex coefficients 
$C_{j,n}$. The time development is obtained by propagating the coupled 
equations of motion 
\cite{zakrzewski2005,schiro2011,wernsdorfer2010,polkovnikov2005,natu2011}
\beq
i \frac{\partial C_{j,n}}{\partial t} = 
\frac{\partial}{\partial C_{j,n}^*} \langle \psi_G(t)| H |\psi_G(t)\rangle. 
\enq 
In order to diagnose the state, the local total density $n_j$ and condensate 
wave function $\Psi_j=\langle \ah_j\rangle$ are computed. We will 
characterize the system by studying the behavior of $n_j$, the 
local condensate density $n_{cj}=|\Psi_j|^2$, and phase $\varphi={\rm arg} \Psi$.
Although simplistic and certainly quantitatively inaccurate in low 
dimensions, I see this mean-field study of the time development as 
a first study which 
can later be vindicated or falsified in more accurate simulation schemes.

The phase diagram for the Bose-Hubbard model was discussed in, e.g., 
Ref.\ \cite{kuhner1998}. The mean-field critical point at $n=1$ 
in 2D is $t\approx 0.042U$ at $\mu=0.5U$. 
If the system is in an external potential, the density is usually well 
modeled by a local-density approximation (LDA), so that alternating 
superfluid and Mott insulating regions exist alongside each other, 
determined by the local chemical potential $\mu-V(r)$. The density 
profile assumes a characteristic wedding-cake structure \cite{bergkvist2004}.

\section{Outflow from a Mott insulator}
\label{sec:gaussian}

\begin{figure}
\includegraphics[width=0.95\columnwidth]{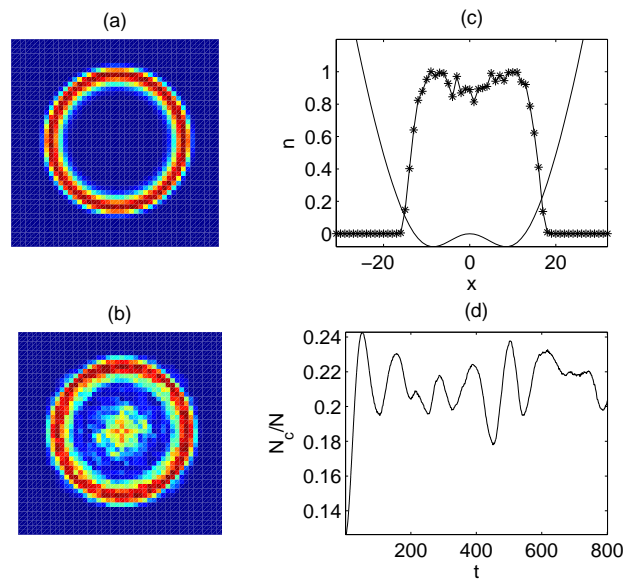} 
\caption{[Color online] 
Time development of a 2D lattice Bose gas following the turn-on 
of a wide Gaussian perturbation potential. 
(a) Spatial distribution of the condensate density before the switch 
of the potential. 
(b) Final condensate density, 
after an evolution of duration $t=800U^{-1}$. 
(c) Cross-section of the final total density profile (symbols) and 
potential profile after the switch (line). 
(b) Fraction of Bose-Einstein condensed atoms as a function of time.
The tunneling matrix element is $J=0.03U$, the trap frequency is 
$\omega=0.07U$, the chemical potential is $\mu=0.5U$, and the 
Gaussian potential has width $W=10$ lattice sites and strength 
$V_1=0.5U$.
}
\label{fig:gauss_weak}
\end{figure}

We consider a 2D system trapped in a harmonic potential, 
\beq
V_{0}(r_j) = \frac{\omega^2}{2}(x_j^2+y_j^2),
\enq
where the integer coordinates $x_j$ and $y_j$ run from $-L+1$ to $L$ as 
$j$ runs from 0 to $4L^2-1$. 
With the choice $J=0.03U$ and $\mu=0.5U$, the system is Mott insulating 
in the center with a surrounding superfluid shell. 
The initial condensate density profile, with $L=32$ and 
$\omega=0.07U$, is graphed in Figure \ref{fig:gauss_weak}(a). 
At time $t=0$, with the system in the ground state of the trap, an 
additional perturbing potential is suddenly switched on, 
\beq
V(r_j) =  V_0(r_j) + V_1 e^{-(x_j^2+y_j^2)/W^2}.
\enq
The resulting total potential assumes a toroidal form, as can be 
seen in Fig.\ \ref{fig:gauss_weak}(c).
In
the case of Fig.\ \ref{fig:gauss_weak} I chose $V_1=0.5U$ and 
$W=10$, to make a relatively wide perturbation. 
There results a rather violent time development, but over time it is 
seen that the Mott insulating phase is molten and the system assumes 
something reminiscent of a steady state; Figs.\ \ref{fig:gauss_weak}(b)-(c) 
indicate that apart from the noise induced (which I would like to 
call thermal noise, although the relation between the current mean-field 
approximation and a finite-temperature Hubbard model is not at all clear) 
the final density follows the potential profile, closely approximating the LDA. 
In Fig.\ \ref{fig:gauss_weak}(d), the total condensate fraction 
$N_c=\sum_j n_{cj}$ 
is plotted as a function of time. As expected it is increasing as the 
Mott insulator melts, and the time scale of the process is of the order 
of $100 U^{-1}$ or $10 J^{-1}$.

\begin{figure}
\includegraphics[width=0.95\columnwidth]{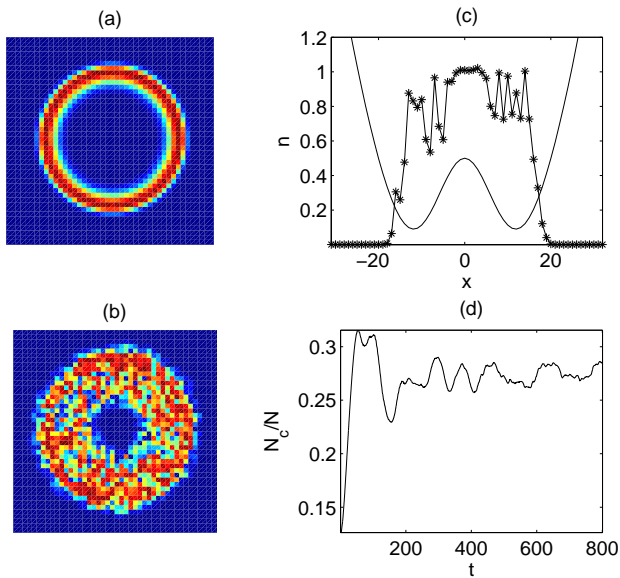} 
\caption{[Color online] 
Time development of a 2D lattice Bose gas following the turn-on 
of a wide Gaussian perturbation potential. 
Panels and parameters are as in Fig.\ \ref{fig:gauss_weak}, 
except the height of the Gaussian potential, $V_1=1.0U$.
}
\label{fig:gauss_strong}
\end{figure}

However, the picture is different if the potential is switched to 
large enough values. In Fig.\ \ref{fig:gauss_strong}, 
the final potential strength is now set to the larger value $V_1=1.0U$. 
In this case, the Mott insulator resists melting and the system 
is hindered from approaching equilibrium. A steady state seems 
to be attained, again after a few inverted hopping periods $J^{-1}$, 
but it is not the LDA distribution seen in the 
previous numerical experiment. 
Figure \ref{fig:gauss_strong}(b-c) show that there is still left a 
Mott insulating region with unit density at the trap center. 
In Figure \ref{fig:gauss_strong}(d) we see that in this case, the 
condensate fraction has also increased a bit, but the important 
finding is that the central Mott insulator is not entirely molten. 
This behavior persists over a range of values of the inverted final 
potential. Further numerical experimentation shows that when 
$V_1$ is larger than about $0.8U$, the Mott insulator insulates. 

\begin{figure}
\includegraphics[width=0.95\columnwidth]{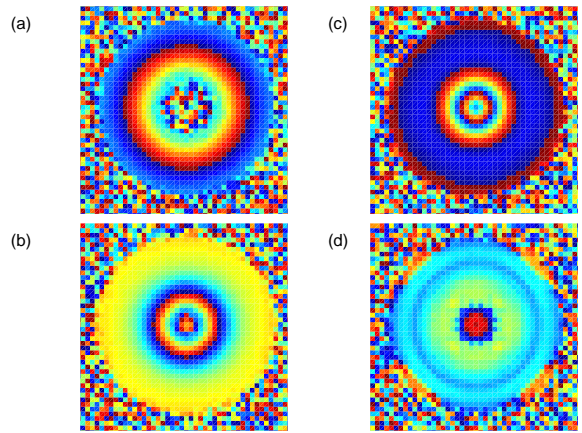} 
\caption{[Color online] 
Spatial distribution of the condensate phase of 
a 2D lattice Bose gas in a harmonic trap plus Gaussian potential. 
Parameters are as in Fig.\ \ref{fig:gauss_weak}, featuring the weaker 
Gaussian potential with $V_1=0.5U$, and the panels 
refer to times (a) $t=20U^{-1}$, (b) $t=40U^{-1}$, (c) $t=60U^{-1}$, 
and (d) $t=100U^{-1}$.
}
\label{fig:phase_weak}
\end{figure}

\begin{figure}
\includegraphics[width=0.95\columnwidth]{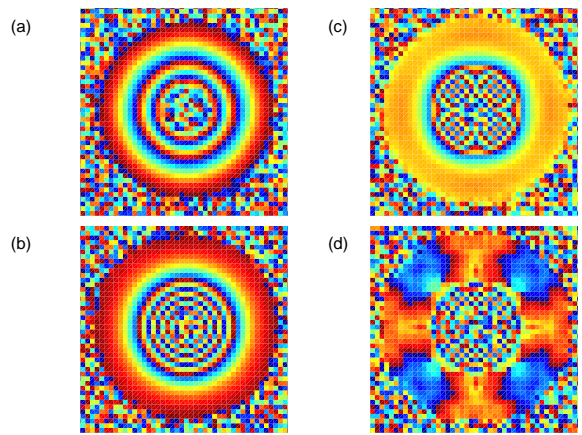} 
\caption{[Color online] 
Spatial distribution of the condensate phase of 
a 2D lattice Bose gas in a harmonic trap plus Gaussian potential. 
Parameters are as in Fig.\ \ref{fig:gauss_strong}, featuring 
the stronger Gaussian potential with $V_1=1.0U$, and the panels 
refer to times (a) $t=20U^{-1}$, (b) $t=40U^{-1}$, (c) $t=60U^{-1}$, 
and (d) $t=100U^{-1}$.
}
\label{fig:phase_strong}
\end{figure}

We have thus found that if the potential is changed a little, 
the Mott insulator will melt, but if it is changed a lot, it 
will not. The reason for this behavior becomes clearer if one 
studies the velocity of the gas. In Fig.\ \ref{fig:phase_weak} the 
phase of the condensate is plotted for the first numerical 
experiment. 
In Fig.\ \ref{fig:phase_strong} it is plotted for the second one. 
The fluid velocity is related to the wavenumber of the phase 
variation pattern through the relation $v_j=J\sin k_j$, 
where $j=x,y$ denotes a Cartesian coordinate axis. The wavenumbers 
$k_j$ are restricted to lie in the interval $(-\pi,\pi)$. 
If the potential is steep, the fluid will be accelerated to 
the maximum wavenumber $\pi$, as is seen to be the case in 
Fig.\ \ref{fig:phase_strong}. This gives rise to a 
current reversal, which is in essence a Bloch oscillation. 
At longer times turbulent processes arise, 
possibly aided by the dynamical instability known to take place 
in a condensate at $k=\pi/2$ \cite{desarlo2005}, and in interacting 
systems at even smaller wavevectors \cite{polkovnikov2005}.
The subsequent 
time development is noisy. This in itself would not necessarily 
stop the fluid from eventually flowing into the newly created potential 
wells, and it does not do so if the system is purely superfluid 
(as seen in simulations not shown here). 
However, when a Mott insulating region is in the way, the simulations 
indicate that the whole process is 
halted and the Mott insulator insulates. 

Snoek and Hofstetter \cite{snoek2007} studied the dynamics of a trapped 
lattice boson system following lateral displacement of the trap. Similarly 
to the present study, they found that Bloch oscillations play a role for the 
dynamics in the case of large displacements. However, in that study, the 
system always relaxed towards the new equilibrium in two dimensions. In the 
present setting, we see that this behavior can be violated. 

\begin{figure}
\includegraphics[width=0.95\columnwidth]{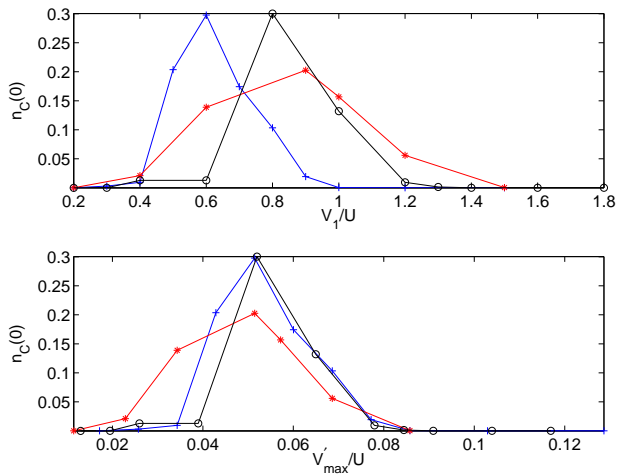} 
\caption{
Final central condensate density after the turn-on 
of a wide Gaussian perturbation potential and subsequent evolution 
for a duration of $800U^{-1}$. The condensate 
density $n_c$ is averaged over the nine centermost points of the 
lattice. 
Plusses denote the result of simulations with a Gaussian perturbation 
with width $W=10$ and height $V_1$; asterisks are for a Gaussian 
with width $W=15$, and circles are for a Lorenzian profile with 
width $W=15$. 
In (a), the x axis is the height of the perturbing potential, $V_1$; 
in (b), it denotes the maximum potential slope, $V'_{\rm max}$.
}
\label{fig:sweepv}
\end{figure}

A wider scan over parameters is collected in Fig.\ \ref{fig:sweepv}. 
Here, we have measured the condensate density $n_c$ in the center at the 
end of the simulation, and plotted it against the potential height 
$V_0$ as well as against the maximum slope $V'_{\rm max}$. Two different 
Gaussian perturbations are chosen, with widths $W=10$ and $W=15$. 
In addition, I tried a perturbation with a Lorenzian profile, not 
because of experimental relevance (it is probably very hard to make), 
but in order to show that the qualitative result is insensitive to 
the shape of the potential. It is seen that the core first becomes 
superfluid when the potential height $V_1$ exceeds about 0.4$U$, simply because 
of LDA considerations. Then, for large enough values of $V_1$, there 
is no superfluid in the core, as we have seen above. This cutoff 
value depends on the specific potential, as seen in 
Fig.\ \ref{fig:sweepv}(a), but in Fig.\ \ref{fig:sweepv}(b), 
the curves collapse onto 
each other when the condensate density is instead plotted against 
the maximum potential slope, $V'_{\rm max}$, i.e., the maximum potential 
difference between adjacent sites. The critical value of 
$V'_{\rm max}$ is in this case approximately 0.08$U$, 
not an integer multiple of $U$, $J$, nor $\omega$, ruling out the 
simplest guesses for resonance physics, but consistent with the picture 
that accelerated atoms experience Bloch oscillations. 

\begin{figure}
\includegraphics[width=0.95\columnwidth]{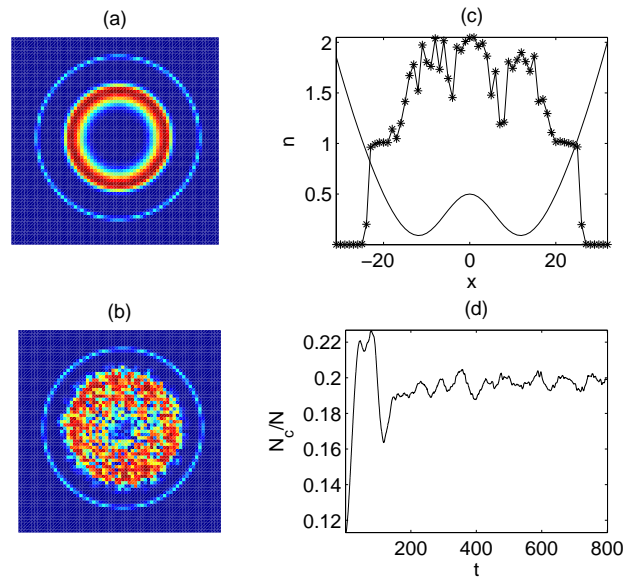} 
\caption{[Color online] 
Time development of a 2D lattice Bose gas following the turn-on 
of a wide Gaussian perturbation potential. 
Panels and parameters are as in Fig.\ \ref{fig:gauss_weak}, 
except $J=0.02U$, $\mu=1.5U$, and 
the height of the Gaussian potential is $V_1=1.0U$.
}
\label{fig:weddingcake}
\end{figure}

So far we have only studied a single combination of tunneling strength 
$J$, trap frequency 
$\omega$ and chemical potential $\mu$, giving a Mott insulator with 
$n=1$. 
Starting in a different Mott insulating region gives the same 
result, as seen in Fig.\ \ref{fig:weddingcake}. 
Here, the parameters are chosen as $J=0.02U$, $\mu=1.5U$, 
and $V_1=1.0U$. This means that the central region in the 
initial state is a $n=2$ Mott insulator surrounded by three 
shells of alternating superfluid and Mott insulator. 
Numerical experimentation 
shows the same behavior as reported above: For a weak enough 
potential, the central Mott insulator is molten, but for a stronger 
one it is not.

\begin{figure}
\includegraphics[width=0.95\columnwidth]{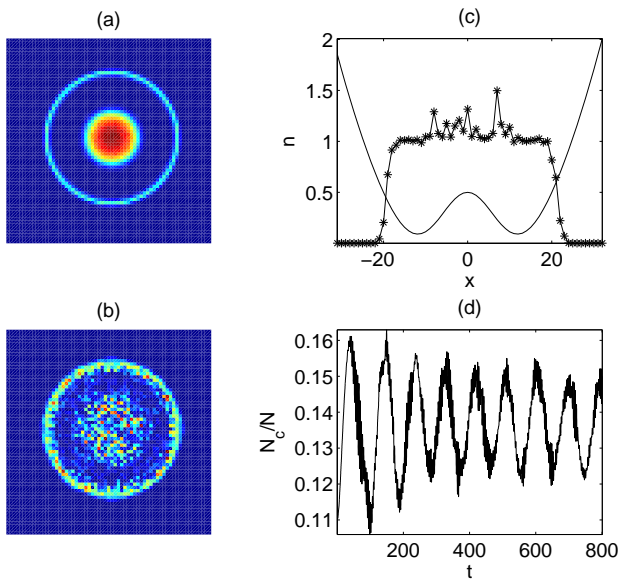} 
\caption{[Color online] 
Time development of a 2D lattice Bose gas following the turn-on 
of a wide Gaussian perturbation potential. 
Panels and parameters are as in Fig.\ \ref{fig:gauss_weak}, 
except $J=0.02U$, $\mu=1.0U$, and 
the height of the Gaussian potential is $V_1=1.5U$.
}
\label{fig:superfluidcore}
\end{figure}

\begin{figure}
\includegraphics[width=0.95\columnwidth]{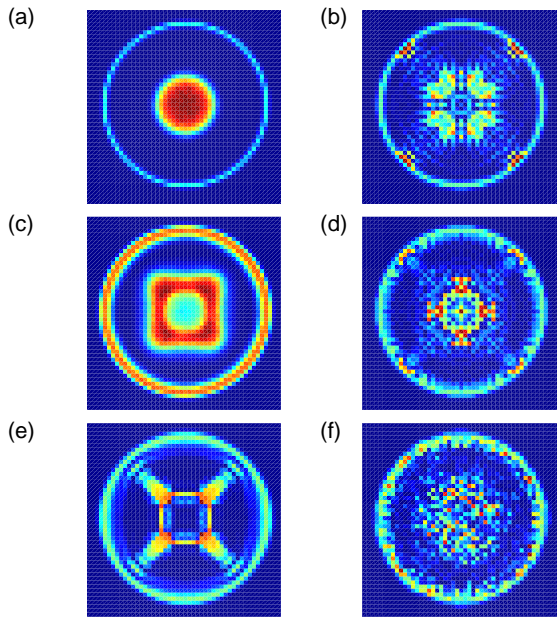} 
\caption{[Color online] 
Spatial distribution of the condensate density $n_C$ of 
a 2D lattice Bose gas in a harmonic trap plus Gaussian potential. 
Parameters are as in Fig.\ \ref{fig:superfluidcore}, featuring 
several superfluid and Mott insulating shells, and the panels 
refer to times (a) $t=0U^{-1}$, (b) $t=30U^{-1}$, (c) $t=60U^{-1}$, 
(d) $t=90U^{-1}$, 
(e) $t=120U^{-1}$, 
and (f) $t=600U^{-1}$.
}
\label{fig:sfcore_time}
\end{figure}

More curious behavior is seen in a system with a superfluid
region in the center. We choose $J=0.3U$, $\mu=1.0U$, and 
$V_1=1.5U$ to produce Fig.\ \ref{fig:superfluidcore}. 
As the peaked potential rises in the center, bosons are flowing 
out from the central superfluid and finally into the outermost 
superfluid shell. However, a partly Mott insulating region is left 
in the center, with a few superfluid atoms intermixed, and then, subsequent dynamics is halted. It is 
seen in Fig.\ \ref{fig:superfluidcore} that the final steady 
state does not rhyme very well with the applied potential. 
A closer look at the time dependence is given in Fig.\ 
\ref{fig:sfcore_time}. 
It is seen that the superfluid atoms flow out by creating four jets 
through the Mott insulator, following the spatial symmetry of the 
lattice. A noisy state is left behind, containing a small amount 
of condensate, but as seen above, the final steady state clearly 
violates the LDA. 
In this case, it appears that the outflowing superfluid bosons leave a Mott 
insulator behind; however, a quantitative explanation seems to be out 
of reach at the present.

\section{Stirring a Mott insulator}
\label{sec:spoon}
Now let us investigate what it takes to make a hole 
in a Mott insulator. To this end, we take a system with 
$\mu=0.5$ as above; as we have seen this gives a Mott insulator 
in the center with an approximate width of 20 lattice sites.
We move a narrow Gaussian ``spoon'' potential through the 
cloud, as was done in Ref.\ \cite{raman2001} to excite 
vortex pairs in a BEC (cf.\ \cite{lundh2003}). 
To produce a reasonably narrow and strong stirrer I choose the parameters 
$W=3$ and $V_1=5.0U$, and the center of the potential 
moves linearly through the cloud with a velocity $v$, 
\beq
V(r_j,t) =  V_0(r_j) + V_1 e^{-((x_j-vt)^2+y_j^2)/W^2}.
\enq 
First, we conclude that a Mott insulator will not be perturbed 
unless the spoon starts in a superfluid region. This is in 
accordance with the definition of Mott insulator, and a series of 
numerical experiments (not shown here) have confirmed this. 
Thus, we start the spoon at the edge of the simulation cell and 
move it through the superfluid shell and then to the Mott 
insulating interior of the cloud. 

\begin{figure}
\includegraphics[width=0.95\columnwidth]{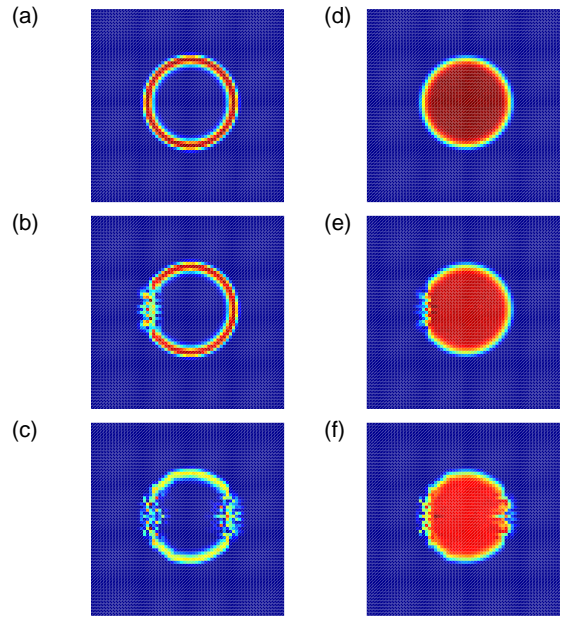} 
\caption{[Color online] 
Time development of a 2D lattice Bose gas as a 
Gaussian ``spoon'' potential is moved through it. 
The potential has width $W=3$, amplitude $V_1=5U$, 
and velocity $v=0.5U$. The remaining parameters
are as in Fig.\ \ref{fig:gauss_weak}.
}
\label{fig:spoon_fast}
\end{figure}

\begin{figure}
\includegraphics[width=0.95\columnwidth]{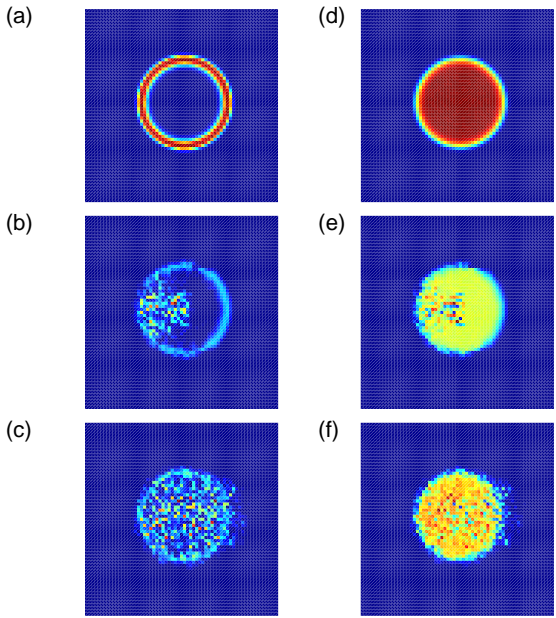} 
\caption{[Color online] 
As in Fig.\ \ref{fig:spoon_fast}, but with 
velocity $v=0.025U$. 
}
\label{fig:spoon_slow}
\end{figure}

\begin{figure}
\includegraphics[width=0.95\columnwidth]{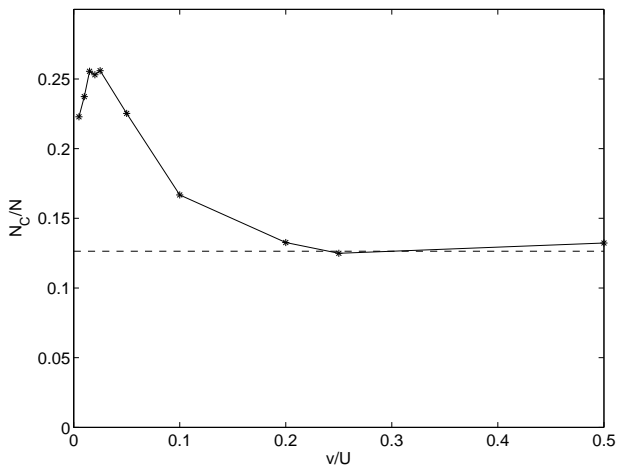} 
\caption{Fraction of Bose-Einstein condensed atoms at the end 
of a sweep with a Gaussian ``spoon'' potential, as a function 
of the velocity with which the spoon is moved. 
Dashed line: Initial condensate fraction.
}
\label{fig:spoon_collect}
\end{figure}

Figure \ref{fig:spoon_fast} shows what happens when the spoon 
is moved quickly through the cloud: The Mott insulator insulates, 
and the presence of the spoon is only felt as it passes through 
the thin superfluid shell. 
In this simulation we chose a velocity $v=0.5U$
(recall that $\hbar=1$ and 
the lattice constant is 1). This is, 
in fact, too fast for the tunneling dynamics to respond; in 
order to melt the insulator one needs to move across a lattice site at 
a time scale comparable with $J^{-1}$. Such a case is shown in 
Figure \ref{fig:spoon_slow}, where $v=0.025U$. 
Here we see how the spoon, given enough time, excites 
bosons out of the Mott insulator and depletes the density 
in the vicinity. An attempt to quantify this 
result is made in Fig.\ \ref{fig:spoon_collect}. 
Here, we measure the fraction of condensed atoms $N_c/N$ 
at the final time, when the spoon has traversed 
the entire simulation cell with the length of 64 sites. 
It is seen that the upper critical velocity for exciting 
atoms into the condensate is around $v=0.1U$, which is 
comparable to the tunneling $J$. We note that the 
sound velocity is likewise of order $J$. These 
simulation results indicate that making a hole in the Mott 
insulator is basically a question of moving slow enough, 
or else the Mott insulator will not budge. When the velocity 
is decreased below the tunneling strength, though, 
the curve is seen to decrease again. Indeed, it is natural to 
expect that for a slow enough spoon, the fluid will adjust 
adiabatically and the perturbation will again be minimized. 
Thus, in the case of a moving spoon, we have identified 
three parameter regimes: An adiabatic regime for velocities 
about an order of magnitude below $J$; a supersonic regime 
in which the edge superfluid has no time to react; and 
an intermediate regime in which the spoon is slow enough to 
excite the Mott insulating atoms and fast enough to do 
maximum damage.

\section{Conclusion}
\label{sec:conclusion}

In summary, I have studied the hydrodynamic behavior 
of a system of lattice bosons in an external potential, 
by means of a number of numerical experiments. 
In the first set of 
experiments, the time development 
is monitored after a sudden switch of the potential. 
It is found that if the potential is switched by a small enough 
amount, the bosons will move in order to adjust to the new potential 
and a Mott insulator may melt. However, if the switch is large enough, 
the Mott insulator will not give in and the system stays in a 
metastable configuration. The reason for this behavior is that 
the supercurrent breaks down in a Bloch oscillation if the potential 
switch is too large. In the second set of numerical experiments, 
a ``spoon''-type of localized potential is moved through a system 
containing a Mott insulating region. The Mott insulator is affected 
by the spoon if is moved at just the right pace, so that the time scale 
for traversing a lattice site is comparable to the tunneling 
time scale, $J^{-1}$. For faster spoons, its presence is not 
felt in the Mott insulator, and for slower spoons, the whole process 
is adiabatic. 

Experimentally, present techniques for in-situ measurement of 
filling factor \cite{folling2006,scherson2010} is the most straightforward 
way of testing the predictions made here. 
A classic time-of-flight measurement, which in effect can tell the 
Bose-Einstein condensed fraction of atoms \cite{greiner2002}, 
is also feasible, as can 
be seen from the plots of said quantity in the figures.

\begin{acknowledgments}
This work was supported by the Swedish Research Council, 
(Vetenskapsr{\aa}det), and was conducted using the resources of High 
Performance Computing Center North (HPC2N).
\end{acknowledgments}


\end{document}